\documentclass[11pt,twoside]{article}

\usepackage{asp2004}
\usepackage{epsf}
\usepackage{psfig}
\usepackage{lscape}

\begin{document}
\title{From ISO to Spitzer: a new view of the AGB-PN transition phase}    
\author{P. Garc\'\i a-Lario$^1$, J.V. Perea Calder\'on$^2$}   
\affil{$^1$ESA/European Space Astronomy Centre. ISO Data Centre, E-28080 
Madrid, Spain\\
$^2$European Space Astronomy Centre. ISO Data Centre, Madrid, Spain}    

\begin{abstract} 
 We present a  novel classification
scheme for stars evolving in the transition phase between the Asymptotic 
Giant Branch (AGB) to the Planetary Nebula (PN) stage based on the results
obtained with ISO-SWS.  With the better sensitivity and higher
spatial resolution of Spitzer this analysis 
can now be extended to  a larger number of sources located in 
the Galactic Bulge, in the Magellanic Clouds and in other 
Galaxies of the Local Group,  offering an excellent opportunity to study 
the validity of the proposed scheme in environments which are free of the
distance scale bias that hinder the observations made with ISO on galactic
sources located at uncertain distances. The new observations
will be used to test the current evolutionary  models  which 
predict the dredge-up of processed material to the surface of  low- and 
intermediate-mass stars as a function of the progenitor mass and  of the 
metallicity.
\end{abstract}


\section{A novel classification scheme based on ISO results}

Based on the analysis of $\sim$350 ISO SWS spectra of sources in the AGB-PN 
transition phase retrieved from the ISO Data Archive we have been able to 
propose a novel classification scheme which takes into account the evolution 
of the overall shape of the infrared spectrum in combination with the
gas-phase molecular bands and solid state features detected in the SWS 
spectral range. We identify three main chemical evolutionary branches which 
are interpreted as the result of the evolution of low-mass ($<$1.5--2.0 
M$_{\odot}$), intermediate-mass (2.0--3.0 M$_{\odot}$) and high-mass ($>$ 
3.0--4.0 M$_{\odot}$) AGB stars, respectively.

 The sequences reflect the increase of optical thickness in the 
circumstellar shell of AGB stars  followed by its cool down as a consequence 
of the shell expansion after the end of the strong mass loss phase. It is 
also a sequence in which we can follow the process of condensation and growth
of the dust grains in the envelope until the star becomes a PN. In addition, 
there is a clear evolution of carbonaceous material from aliphatic  to 
aromatic structures and of the silicates from amorphous to crystalline, 
which is still not very well understood. 

The proposed scheme is consistent with the theoretical model predictions 
based on the dredge-up of processed material to the surface of AGB stars 
which result in the transformation of the initially O-rich AGB star into a
C-rich AGB star, after a few thermal pulses. The number and efficiency of 
the thermal pulses in low-mass stars is expected to be small. As a
consequence of this, low-mass stars would stay  as O-rich  during the whole 
AGB-PN evolution. Only part of them will become low-mass O-rich type II or 
type III PNe. In the most extreme cases very low-mass stars will not develop 
an observable PN. Intermediate-mass stars  will 
soon become C-rich and will further evolve as such until the PN stage. 
High-mass stars  will develop very thick envelopes 
and thus strong silicate absorption bands and will
activate the {\it hot bottom burning} mechanism which prevents the 
formation of carbon and favours the production of nitrogen, instead, 
evolving as O-rich stars until they become high-mass, N-rich type I PNe. 

\section{What can Spitzer do that ISO could not do?}


\subsection{More observations...}

  The number of sources in the AGB-PN evolutionary phase observed with ISO 
is small. Some of the new results derived from ISO data will need to be
confirmed by extending the observations to a larger number of sources.  
Spitzer expected lifetime of $\sim$5 yr could provide a huge increase of 
available data to further study the many questions still left open by ISO. 

\subsection{Of fainter sources...}

  Below the 1--5 Jy level, the sensitivity of ISO-SWS was not enough to 
obtain spectra with enough quality to derive reliable conclusions.
 With the much higher sensitivity of Spitzer it will be possible to observe 
under much better conditions these sources and extend the analysis to other 
sources several orders of magnitude fainter.

\subsection{Located at homogeneous distances...}

   The study of the stars in the ISO sample above described is hampered by 
the poor knowledge of their distances. This problem can be overcome with
Spitzer if the analysis concentrates on well-defined samples located at
distances which are known with a reasonable accuracy. This includes the
Galacic Bulge, the Small and Large Magellanic Clouds and other galaxies of 
the Local Group, which were forbidden for ISO (because of its poor 
sensitivity).  As a byproduct, we will be able to analyze also the influence
of metallicity on the proposed evolutionary scheme, expected to play an 
important role according to the existing models.

\subsection{With a higher spatial resolution...}

  It has been suggested that PAHs may be predominantly present in the 
scattering lobes, while the  crystalline silicates are expected to be 
present in the disks of bipolar sources showing a mixed chemistry. With the 
high spatial resolution of Spitzer it will be possible to resolve the
different emitting regions in sources extended 
over just a few arcsec showing this dual chemistry.

\subsection{Extending the search to other galaxies}

 The use of multi-filter photometry with Spitzer (IRS + MIPS) can be a
powerful tool to search for new sources in the AGB-PN evolutionary phase. 
Color-color diagrams based on large-scale maps of galaxies of the Local 
Group in the adequate filters could provide us with large datasets of  new 
candidate sources which may have escaped from detection in classical optical 
surveys. 




\end{document}